%% file: EWK-11-017_temp.tex
\pdfoutput=1

\documentclass[11pt,twoside,a4paper,cmspaper,final,collab]{cms-tdr}
\def\svnVersion{175800}\def\cmsCernNo{2012-226}\def\cmsCernDate{\today}\def\cmsMessage{Submitted to Physical Review Letters}

\begin{document}\cmsNoteHeader{EWK-11-017}

\hyphenation{had-ron-i-za-tion}
\hyphenation{cal-or-i-me-ter}
\hyphenation{de-vices}

\RCS$Revision: 160424 $
\RCS$HeadURL: svn+ssh://svn.cern.ch/reps/tdr2/papers/EWK-11-017/trunk/EWK-11-017.tex $
\RCS$Id: EWK-11-017.tex 160424 2012-12-06 20:22:42Z andersj $

\newcommand{\pb}{\ensuremath{\mathrm{pb}}}%
\newcommand{\pp}{\ensuremath{\Pp\Pp}}%
\newcommand{\rts}{\ensuremath{\sqrt{s}}}%
\newcommand{\ra}{\ensuremath{\rightarrow}}%
\newcommand{\MN}{\Pgm\Pgn}%
\renewcommand{\EE}{\Pep\Pem}%
\newcommand{\EN}{\Pe\Pgn}%
\newcommand{\LN}{\ensuremath{\ell\Pgn}}%
\newcommand{\MT}{\ensuremath{M_{\mathrm{T}}}\xspace}%
\newcommand{\MLL}{\ensuremath{\mathrm{M}_{\ell\ell}}}%
\newcommand{\met}{\ensuremath{{E\!\!\!\!/}_{\mathrm{T}}}\xspace}
\newcommand{\Wmn}{\ensuremath{\PW \ra \MN}}%
\newcommand{\Wpmn}{\ensuremath{\PWp \ra \Pgmp\Pgn}}%
\newcommand{\Wmmn}{\ensuremath{\PWm \ra \Pgmm\Pagn}}%
\newcommand{\Zmm}{\ensuremath{\PZz \ra \MM}}%
\newcommand{\Wtn}{\ensuremath{\PW \ra \Pgt\Pgn}}%
\newcommand{\Ztt}{\ensuremath{\PZz \ra \Pgt^+\Pgt^-}}%
\newcommand{\ppZmm}{\pp \ra \PZz(\gamma^*) + X \ra \MM + X}%
\newcommand{\ppWmn}{\pp \ra \PW + X \ra \MN + X}%
\newcommand{\ppWpmn}{\pp \ra \PWp + X \ra \Pgmp\Pgn + X}%
\newcommand{\ppWmmn}{\pp \ra \PWm + X \ra \Pgmm\Pagn + X}%
\newcommand{\Wen}{\ensuremath{\PW \ra \EN}}%
\newcommand{\Wpen}{\ensuremath{\PWp \ra \Pep\Pgn}}%
\newcommand{\Wmen}{\ensuremath{\PWm \ra \Pem\Pagn}}%
\newcommand{\Zee}{\ensuremath{\PZz \ra \EE}}%
\newcommand{\ppZee}{\pp \ra \PZz(\gamma^*) + X \ra \EE + X}%
\newcommand{\ppWen}{\pp \ra \PW + X \ra \EN + X}%
\newcommand{\ppWpen}{\pp \ra \PWp + X \ra \Pep\nu  + X}%
\newcommand{\ppWmen}{\pp \ra \PWm + X \ra \Pem\Pagn + X}%

\newcommand{\Wln}{\ensuremath{\PW \ra \LN}}%
\newcommand{\Wpln}{\ensuremath{\PWp \ra \ell^+\Pgn}}%
\newcommand{\Wmln}{\ensuremath{\PWm \ra \ell^-\Pagn}}%
\newcommand{\Zll}{\ensuremath{\PZz \ra \ell^+ \ell^-}}%
\newcommand{\ppZll}{\pp \ra \PZz(\gamma^*) + X \ra \ell^+ \ell^- + X}%
\newcommand{\ppWln}{\pp \ra \PW + X \ra \LN + X}%
\newcommand{\ppWpln}{\pp \ra \PWp + X \ra \ell^+\Pgn  + X}%
\newcommand{\ppWmln}{\pp \ra \PWm + X \ra \ell^-\Pagn + X}%

\newcommand{\Wptn}{\ensuremath{\PWp \ra \Pgt^+\Pgngt}}%
\newcommand{\Wmtn}{\ensuremath{\PWm \ra \Pgt^-\Pagn}}%

\newcommand{\Wev}{\Wen}
\newcommand{\Wmv}{\Wmn}
\newcommand{\Ztautau}{\Ztt}
\renewcommand{\MET}{\met}
\newcommand{\gammaZ}{\ensuremath{\gamma^{*}\PZz}}
\newcommand{\gammaZmm}{\mbox{$ \gamma^{*}/\PZz\rightarrow \MM$}}
\newcommand{\gammaZee}{\mbox{$ \gamma^{*}\PZz\rightarrow \Pep  \Pem$}}
\newcommand{\gammaZtt}{\mbox{$ \gamma^{*}\PZz\rightarrow \Pgt^{+}  \Pgt^{-}$}}
\newcommand{\gammaZll}{\mbox{$ \gamma^{*}\PZz\rightarrow \ell^{+}  \ell^{-}$}}
\newcommand{\invpb}{\mbox{$\text{pb}^{-1}$}}
\newcommand{\invnb}{\mbox{$\text{nb}^{-1}$}}
\newcommand{\mz}{\mbox{$ m_{\cPZ}$}}
\newcommand{\hta}{\mbox{$ \eta$}}
\newcommand{\fh}{\mbox{$ \phi$}}
\newcommand{\etot}{\mbox{$ \epsilon_\text{tot}$}}
\newcommand{\eclustering}{\mbox{$ \epsilon_\text{clustering}$}}
\newcommand{\etracking}{\mbox{$ \epsilon_\text{tracking}$}}
\newcommand{\egsfele}{\mbox{$ \epsilon_\text{gsfele}$}}
\newcommand{\epreselection}{\mbox{$ \epsilon_\text{preselection}$}}
\newcommand{\eisolation}{\mbox{$ \epsilon_\text{isolation}$}}
\newcommand{\eclassification}{\mbox{$ \epsilon_\text{classification}$}}
\newcommand{\eelID}{\mbox{$ \epsilon_\text{elID}$}}
\newcommand{\etrigger}{\mbox{$ \epsilon_\text{trigger}$}}

\newcommand{\DE}{$\Delta\eta_\text{in}$}
\newcommand{\DP}{$\Delta\phi_\text{in}$}
\newcommand{\SEE}{$\sigma_{\eta\eta}$~}
\newcommand{\SEP}{$\sigma_{\eta\phi}$}
\newcommand{\SPP}{$\sigma_{\phi\phi}$}
\newcommand{\SXY}{$\sigma_{XY}$}
\newcommand{\pth}{\hat{p}_{\perp}}
\newcommand{\Lint}{\ensuremath{\mathcal{L}_{\text{int}}}}
\newcommand{\IRelComb} {I^{\text{rel}}_{\text{comb}}}%
\newcommand{\ITRK}     {I_{\text{trk}}}%
\newcommand{\IECAL}    {I_{\text{ECAL}}}%
\newcommand{\IHCAL}    {I_{\text{HCAL}}}%
\newcommand{\Nbg}{N_{\mathrm{bg}}}%

\newcommand{\mumu}{\mu\mu}
\newcommand{\Zmumu}{Z_{\mumu}}
\newcommand{\Zmus}{Z_{\mu s}}
\newcommand{\Zmut}{Z_{\mu t}}
\newcommand{\nonIso}{\text{non\,iso}}
\newcommand{\ZmumuNonIso}{\Zmumu^\nonIso}
\newcommand{\ZmumuTwoHlt}{\Zmumu^{2\mathrm{HLT}}}
\newcommand{\ZmumuOneHlt}{\Zmumu^{1\mathrm{HLT}}}

\newcommand{\NZtomumu}{N_{\Zmm}}

\newcommand{\Nmumu}{N_{\mumu}}
\newcommand{\Nmus}{N_{\mu s}}
\newcommand{\Nmut}{N_{\mu t}}
\newcommand{\NmumuNonIso}{\Nmumu^\nonIso}
\newcommand{\NmumuTwoHlt}{\Nmumu^{2\mathrm{HLT}}}
\newcommand{\NmumuOneHlt}{\Nmumu^{1\mathrm{HLT}}}

\newcommand{\effHlt}{\epsilon_\mathrm{HLT}}
\newcommand{\effIso}{\epsilon_\mathrm{iso}}
\newcommand{\effTrk}{\epsilon_\mathrm{trk}}
\newcommand{\effSa}{\epsilon_{\mathrm{sa}}}

\newcommand{\cls}{CL_{S}}
\newcommand{\mjj}{\ensuremath{m_{jj}}\xspace}
\newcommand{\Wpj}{\PW\ plus jets }

\newlength{\columnfigure}
\newlength{\tripleColFig}
\ifthenelse{\not{\boolean{cms@external}}}{%
  \setlength{\columnfigure}{0.65\columnwidth}
  \setlength{\tripleColFig}{0.48\columnwidth}
}
{%
  \setlength{\columnfigure}{0.9\columnwidth}
  \setlength{\tripleColFig}{0.68\columnwidth}
  \providecommand{\linenomath}{\relax} 
}
\ifthenelse{\boolean{cms@external}}{%
\newcommand{\scotchrule[1]}{\centering\begin{ruledtabular}\begin{tabular}{#1}}
\newcommand{\donescotchrule}{\end{tabular}\end{ruledtabular}}
}{
\newcommand{\scotchrule[1]}{\centering\begin{tabular}{#1}\hline\hline}
\newcommand{\donescotchrule}{\hline\hline\end{tabular}}
}

\cmsNoteHeader{EWK-11-017} 
\title{Study of the dijet mass spectrum in \texorpdfstring{$\Pp\Pp\to\PW+\text{jets}$ events at $\sqrt{s}=7\TeV$}{pp to W+jets events at sqrt(s)=7 TeV}}

\date{\today}

\abstract{ We report an investigation of the invariant mass spectrum
  of the two jets with highest transverse momentum in $\Pp\Pp \to
  \PW$+2-jet and \PW+3-jet events to look for resonant enhancement.
  The data sample corresponds to an integrated luminosity of 5.0\fbinv
  collected with the CMS detector at $\sqrt{s} = 7\TeV$. We find no
  evidence for the anomalous structure reported by the CDF
  Collaboration, and establish an upper limit of 5.0\unit{pb} at 95\%
  confidence level on the production cross section for a generic
  Gaussian signal with mass near 150\GeV.  Additionally, we exclude
  two theoretical models that predict a CDF-like dijet resonance near
  150\GeV.  }

\hypersetup{%
pdfauthor={CMS Collaboration},%
pdftitle={Study of the dijet mass spectrum in pp to W + jets events at sqrt(s)=7 TeV},%
pdfsubject={CMS},%
pdfkeywords={CMS, physics, dijets}}

\maketitle 

The CDF Collaboration reported evidence for an excess in the mass
range 120--160\GeV in the invariant mass (\mjj) spectrum of the two
leading transverse-momentum (\pt) jets produced in
$\Pp\Pap\rightarrow\PW$+2-jet events with a cross section of 
4\unit{pb}~\cite{CDFresult}.  The D\O\ Collaboration carried out a
similar analysis but did not confirm the CDF result, instead setting a
95\% confidence level (CL) upper limit of 1.9\unit{pb} on the cross
section~\cite{D0result}.  This Letter details the search for a
bump-like enhancement in the \mjj spectrum in events with a \PW\ boson
using $5.0\fbinv$ of data collected from pp collisions at $\sqrt{s} =
7\TeV$ with the Compact Muon Solenoid (CMS) detector at the CERN Large
Hadron Collider (LHC) during 2010 and 2011.

We search for a resonance with a width consistent with detector
resolution as reported by CDF.  We further investigate three
representative models, a technicolor $\pi_\mathrm{T}$ from the decay
of a technicolor $\rho_\mathrm{T}$~\cite{Eichten:2011sh}, a
leptophobic $\zp$ decaying to two jets~\cite{Buckley:2011vc}, and the
standard model (SM) Higgs boson ($m_{\PH}=150\,\GeV$) produced in
association with a \PW\ boson (referred to as \PW\PH\ production) and
decaying to a pair of jets.  For the unknown state with detector
resolution, we follow the convention used at the Tevatron of using the
conservative \PW\PH\ simulation for analysis-dependent quantities like
efficiencies and acceptances.  The \PW\PH\ production cross section at
the LHC is negligible compared to contributions from other SM
processes, which overwhelm any contribution to this analysis from
$\PW\PH\rightarrow \ell\nu jj$ decays for $m_{\PH}\approx
125\GeV$~\cite{CMSHiggs,ATLASHiggs}.

A detailed description of the CMS experiment can be found in
Ref.~\cite{CMSDetector}.  The central feature of the CMS detector is a
superconducting solenoid, of 6\unit{m} internal diameter, that
produces an axial magnetic field of 3.8\unit{T}. Located within the
field volume is the silicon pixel and strip tracker extending up to
$|\eta|=2.5$, as well as a lead tungstate crystal electromagnetic
calorimeter (ECAL) and a brass/scintillator hadronic calorimeter
(HCAL), both extending up to $|\eta| =3$.  Outside the field volume in
the forward region ($3 < |\eta| < 5$) is an iron/quartz-fiber hadronic
calorimeter.  Muons are measured in gas-ionization detectors embedded
in the steel return yoke outside the solenoid, in the pseudorapidity
range $|\eta| < 2.4$.  The CMS coordinate system has its origin at the
center of the detector, with the $z$ axis pointing along the direction
of the counterclockwise proton beam. The azimuthal angle is denoted as
$\phi$, the polar angle as $\theta$, and the pseudorapidity is defined
as $\eta=-\ln\left[\tan\left(\theta/2\right)\right]$.

We employ selection criteria similar to those used at the
Tevatron~\cite{CDFresult,D0result}, but modified to adapt to the
higher background rates and different experimental conditions at the
LHC.  We also place more stringent requirements on the jet kinematics,
as suggested in Ref.~\cite{ELM}, to enhance a signal compared to the
irreducible \Wpj background.

Events are selected with one well-identified and isolated lepton (muon
or electron), large missing transverse energy \met, and exactly two or
exactly three high-\pt jets.  The data were collected with a suite of
single-lepton triggers, mostly with a \pt threshold of 24\GeV for
muons and 25--32\GeV for electrons. The trigger efficiency for the
selected muons (electrons) is about 94\% (90\%).  We reconstruct muon
candidates in the region $|\eta| < 2.1$ by combining information from
the silicon tracker and the muon detectors by means of a global fit.
We identify electron candidates within $|\eta|<1.44$ and
$1.57<|\eta|<2.5$ as clustered energy deposits in the electromagnetic
calorimeter that are matched to tracks.  Muon and electron candidates
need to fulfill quality criteria established for the measurement of
the inclusive \PW\ and \cPZ\ cross sections~\cite{WZCMS:2010}.  In
addition, all leptons must be well-separated from hadronic activity in
the event.  Jets within an $\eta$-$\phi$ cone of radius 0.3 around a
lepton candidate are removed.

The muon (electron) transverse momentum must exceed 25~(35)\GeV, and
\met must be greater than 25~(30)\GeV in the muon (electron) analysis.
The transverse mass $\MT$ of each \PW\ candidate must be greater than
50\GeV, where
\begin{linenomath}
\begin{align}
\MT \equiv { \sqrt{2\pt^\ell\,\met\,[1-\cos(\phi_\ell-\phi_{\met})]} }
\nonumber
\end{align}
\end{linenomath}
and $\phi_\ell$ and $\phi_{\met}$ are the azimuthal angles of the lepton
and $\met$, respectively. Events with more than one identified lepton
are vetoed.

We reconstruct jets and \met~\cite{Chatrchyan:2011tn,WZCMS:2010} with
the particle-flow algorithm~\cite{PFT-09-001}, which combines
information from several subdetectors.  The jet finding uses the
anti-$k_\mathrm{T}$ clustering algorithm~\cite{ref:antikt} with a
distance parameter of 0.5.  We require $|\eta_{\textrm{jet}}| < 2.4$
to ensure that they lie within the tracker acceptance, and a minimum
jet \pt of 30\GeV.  Jets must satisfy identification criteria that
eliminate jet candidates originating from noisy channels in the hadron
calorimeter~\cite{Chatrchyan:2009hy}.  Jet-energy corrections are
applied to account for the non-linear response of the calorimeters to
the particle energies and other instrumental effects. These
corrections are based on in-situ measurements using dijet,
$\gamma+{\rm jet}$, and Z+jet data
samples~\cite{Chatrchyan:2011ds}. Overlapping minimum-bias events from
other pp collisions (pile-up) and the underlying event can contribute
additional energy to the reconstructed jets.  The median energy
density due to pile-up is evaluated in each event and the
corresponding energy is subtracted from each
jet~\cite{fastjet1}.  In addition, tracks that do not
originate from the primary vertex are not considered for jet
clustering~\cite{Cacciari:2008gn}.  We verify that the procedures
successfully remove the dependence of jet response on the number of
interactions in a single event.  The jet \pt resolution varies from
15\% at $\pt = 40\GeV$ to 6\% at $\pt =
400\GeV$~\cite{Chatrchyan:2011ds}.  We evaluate the mass resolution
$\sigma_{jj}$ for a selected jet pair using simulation and verify it
using hadronic W decays in data.  We find $\sigma_{jj}$ to be 10\% of
\mjj for masses around $150\GeV$.

We require $\|\vec{p}_{\mathrm{T}}^{~j_1} +\vec{p}_{\text{T}}^{~j_2}\|
> 45\GeV$ and $|\Delta\eta(j_1,j_2)| < 1.2$, where the jets are
numbered in order of decreasing \pt.  We retain events with exactly
two or exactly three jets satisfying $\pt > 30\GeV$ and with the
leading jet having $\pt > 40\GeV$ and pointing more than 0.4\unit{rad}
in azimuth from the direction of the \met.  The selected jets and the
lepton from the \PW\ decay must originate from the same primary
vertex.  Additionally, we impose $0.3 < \pt^{~j_2}/\mjj < 0.7$ to take
advantage of the Jacobian nature of resonant dijet production as
observed in simulation studies compared with nonresonant \Wpj
production.

\PW\ production with two or more jets dominates the selected sample.
Smaller contributions come from top-pair and single-top decays,
Drell--Yan events with two or more jets, multijet production, and
\PW\PW\ and \PW\cPZ\ diboson production where one \PW\ decays into
leptons and the other \PW\ or \cPZ\ decays into quarks.

The shapes of the \mjj distributions for background processes are
modeled using samples of simulated events. The
{\MADGRAPH}5~1.3.30~\cite{MADGRAPH} event generator produces
parton-level events with a \PW\ boson and up to four partons on the
basis of matrix-element (ME) calculations.  (The Tevatron experiments
used the \textsc{alpgen} generator~\cite{Mangano:2002ea}.)  The ME--parton shower
matching scale $\mu$ is taken to be 20\GeV~\cite{Hoche:2006ph}, and
the factorization and renormalization scales are set to $q^2 =
M_{\PW}^2 + p_{\mathrm{T},\PW}^2$.  Samples of \ttbar\ and Drell--Yan
events are also generated with \MADGRAPH.  Single-top production is
modeled with \POWHEG~1.0~\cite{POWHEG}.  Multijet and diboson samples
(\PW\PW, \PW\cPZ, \cPZ\cPZ) are generated with
\PYTHIA~6.422~\cite{Sjostrand:2006za}.  \PYTHIA provides the parton
shower simulation in all cases, with parameters of the underlying
event set to the Z2 tune~\cite{PythiaTuneZ2}.  The set of parton
distribution functions used is \textsc{cteq6ll}~\cite{CTEQ}.
A \GEANTfour-based simulation~\cite{GEANT4} of the CMS
detector is used in the production of all Monte Carlo (MC)
samples. Multiple proton-proton interactions within a bunch crossing
are simulated, and the triggers are emulated.  All simulated
events are reconstructed and analyzed with the same software as data.

We generate signal samples for the \PW\PH model using \PYTHIA, with
parameters corresponding a SM Higgs boson with $m_{\PH}=150\,\GeV$.
We use \PYTHIA for technicolor generation as well.  We generate
leptophobic $\zp$ with \MADGRAPH.  The authors of
Refs.~\cite{Eichten:2011sh,Buckley:2011vc} provided values for masses
and other parameters of the technicolor and $\zp$ models that would
best correspond to the signal observed by CDF.

We determine the contributions of the known SM processes to the
observed \mjj spectrum by means of an extended unbinned
maximum-likelihood fit in the range between 40\GeV and 400\GeV.  We
fit separately in four event categories,
$\{\mu,\Pe\}\times\{2\text{-jet},3\text{-jet}\}$, because the
background compositions differ.  The \mjj signal region, 123 to
186\GeV, corresponding to ${\pm} 2\sigma_{jj}$, is excluded from this
fit in order to arrive at an unbiased estimate of a possible resonant
enhancement in this region.

\begin{table}[btp]
  \topcaption{Treatment of background \mjj shapes and normalizations in a fit
    to the data. The background normalizations are constrained within the
    fit to Gaussian distributions with the listed central values and
    widths.}
\label{tab:Table0}
\scotchrule[lcl]
   Process             &    Shape     & Constraint on normalization \\
   \hline
   \Wpj                &    MC/data   & Unconstrained \\
   Diboson             & MC      & $61.2\unit{pb}\pm 10\%$ (NLO)~\cite{Campbell:2011bn}\\
   \ttbar\             &    MC        & $163\unit{pb}\pm 7\%$ (NLO)~\cite{Kidonakis:2010dk}\\
   Single-top          &    MC        & $84.9\unit{pb}\pm 5\%$ (NNLL)~\cite{Kidonakis:2010tc,Kidonakis:2011wy,Kidonakis:2010ux}\\
   Drell--Yan plus jets &    MC        & $3.05\unit{nb}\pm 4.3\%$ (NNLO)~\cite{FEWZ} \\
   Multijet (QCD)      &    data      & \met fit (described in text) \\
 \donescotchrule
\end{table}

Table~\ref{tab:Table0} lists the SM processes included in the fit.
The \Wpj normalization is a free fit parameter because it is by far
the dominant background.  We allow the normalizations of the other
background components to vary within Gaussian constraints around the
central values also listed in Table~\ref{tab:Table0}.  The central
values for all processes except multijet come from
next-to-leading-order (NLO), next-to-next-to-leading-log (NNLL) or
next-to-NLO (NNLO) calculations, and the constraints reflect the
published uncertainties.  We derive templates for the \mjj
distribution for each background from simulation except for the
multijet events, which contribute when jets are misidentified as
leptons.  In a separate fit to events that fail the lepton isolation
requirements, we determine the central value of the multijet
normalization, the constraint on the normalization and the template
for the \mjj distribution~\cite{WZCMS:2010}.  The fit to data
determines the correlations among the various fit parameters.

The default CMS \MADGRAPH sample of the dominant \Wpj background does
not describe well the \mjj spectrum in the \mjj sidebands.  Four
alternative samples of \PW\ events, with the scales $\mu$ and $q$
increased and reduced by a factor two with respect to those of the
default, fail to provide significant improvement.  Thus, we employ an
empirically-driven combination of three shapes to describe this
component in the fit model:
\begin{linenomath}
\begin{align}
F_{\PW+\text{jets}} = \alpha\, & \mathcal{F}_{\PW+\text{jets}} (\mu_{0}^2, q'^2) +
\beta\, \mathcal{F}_{\PW+\text{jets}} (\mu'^2, q_{0}^2) \nonumber \\
&+ (1-\alpha-\beta)\, \mathcal{F}_{\PW+\text{jets}} (\mu_{0}^2, q_{0}^2)\,,
\label{eqWpjetsShape}
\nonumber
\end{align}
\end{linenomath}
where $\mathcal{F}_{\PW+\text{jets}}$ denotes the \mjj shape from
simulation.  The parameters $\mu_0$ ($\mu'$) and $q_0$ ($q'$)
correspond to the default (alternative) values of $\mu$ and $q$,
respectively, while fractional contributions $\alpha$ and $\beta$ are
free to vary between 0 and 1.  We take $\mu' = 2 \mu_0$ or $0.5 \mu_0$
($q' = 2 q_0$ or $0.5 q_0$), depending on which alternative sample
provides a better fit to data.  Furthermore, we verify, via
pseudo-experiment simulations generated with an alternate shape, that
the function in the above equation has sufficient freedom to describe
the \Wpj shape.

\begin{figure*}[btp]
    \begin{center}

      \includegraphics[width=\tripleColFig]{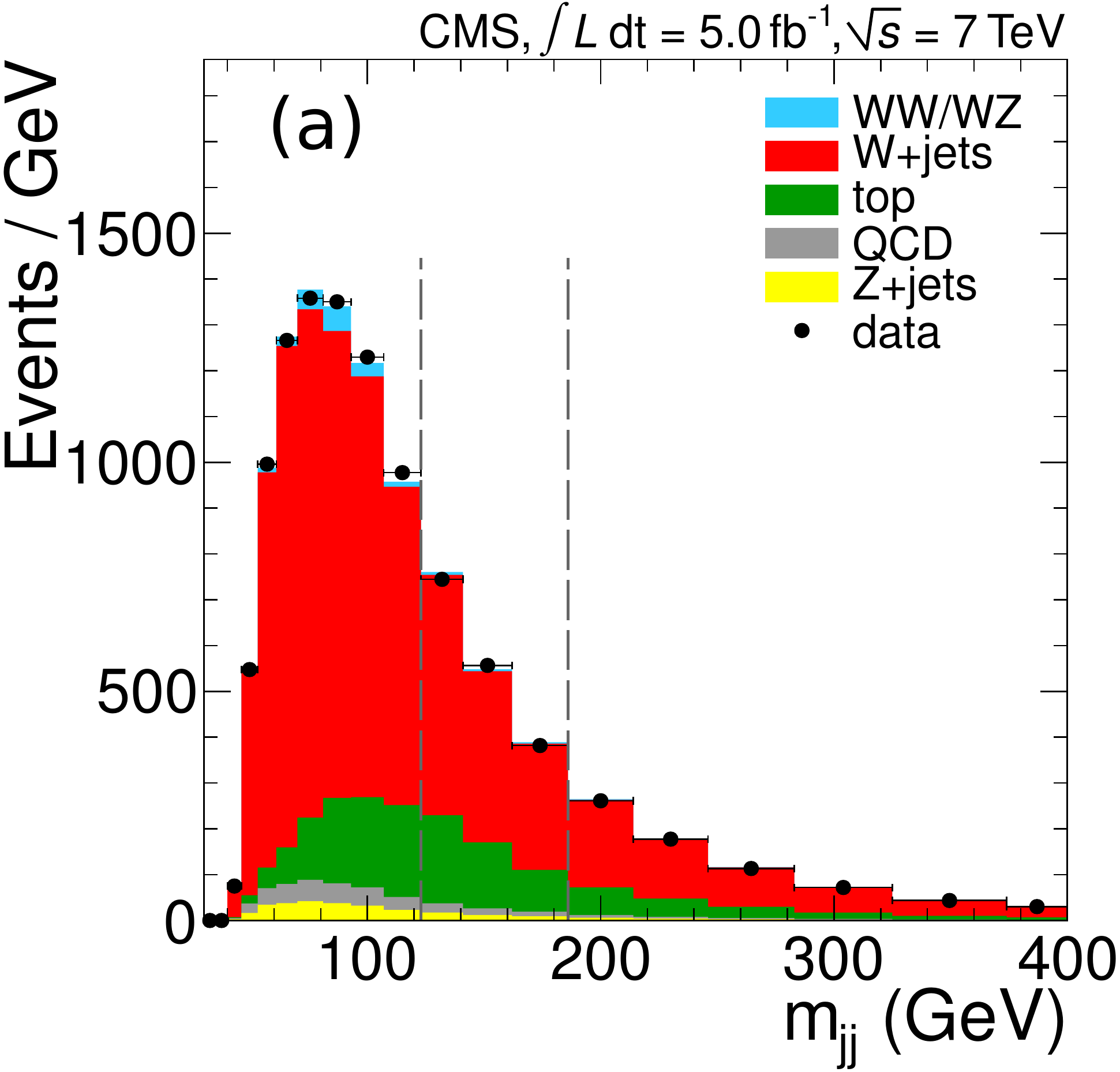}
      \includegraphics[width=\tripleColFig]{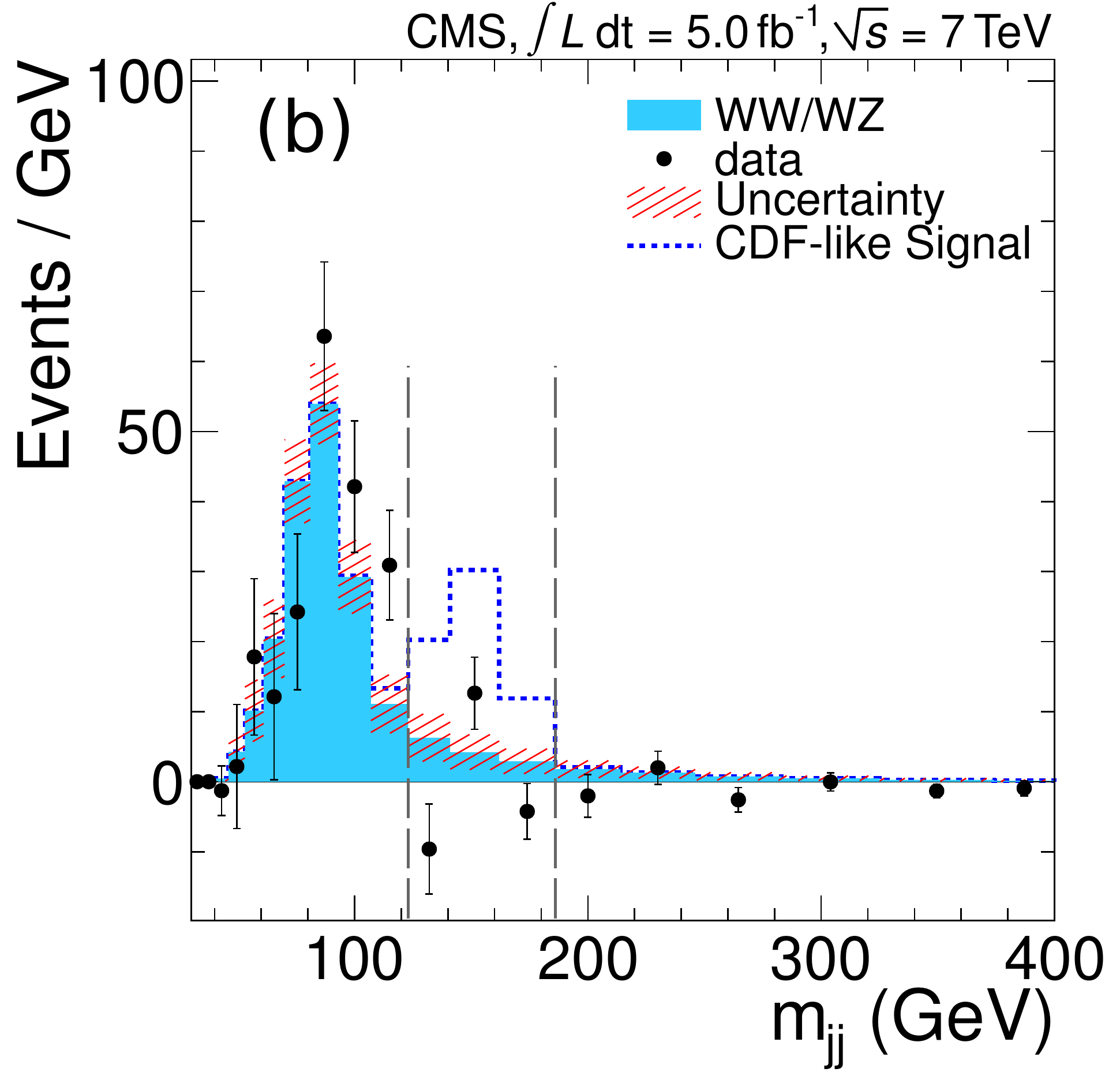}
      \includegraphics[width=\tripleColFig]{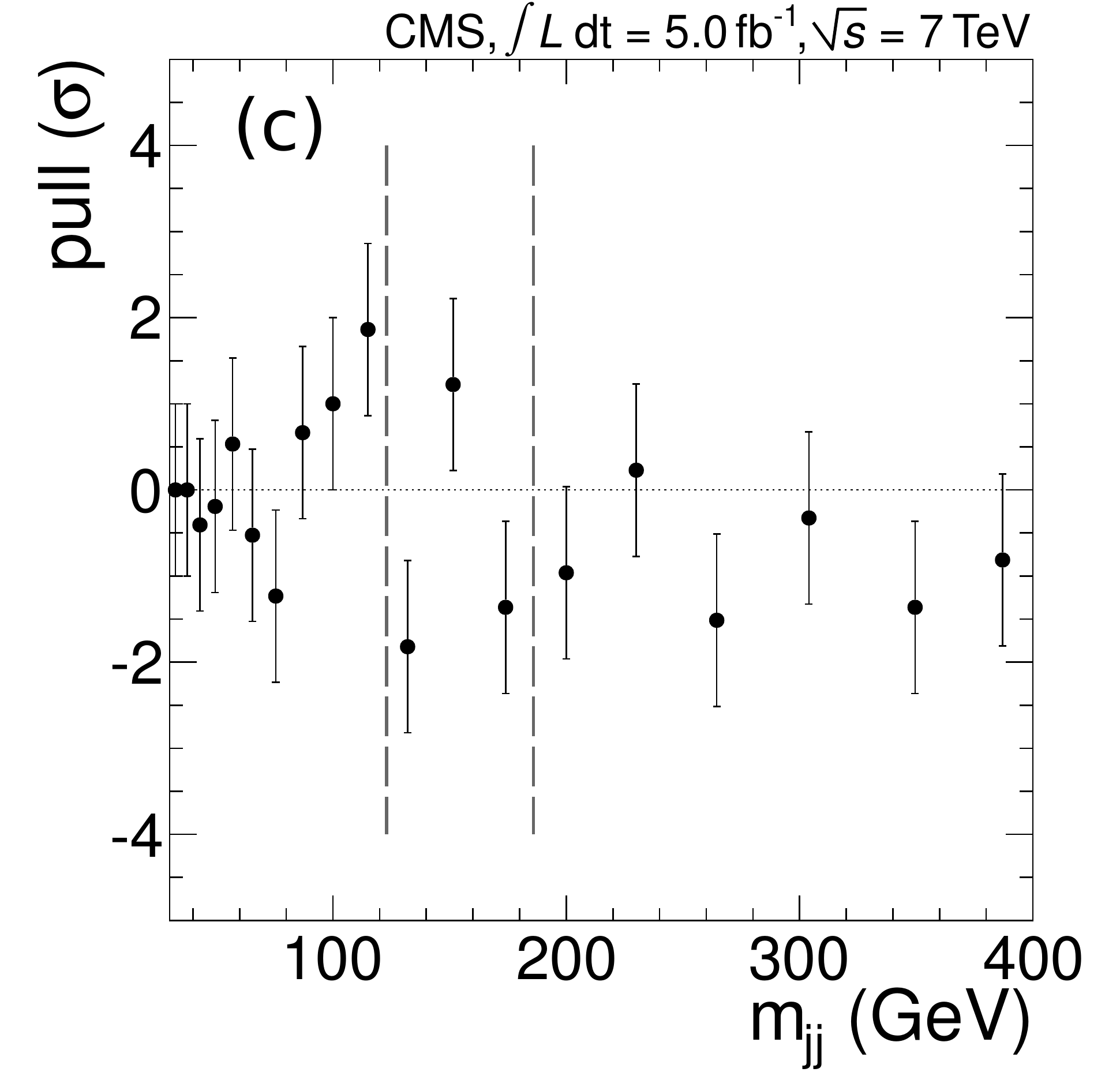}
    \end{center}
    \caption{(a) Distribution of the invariant mass spectrum of the
      leading two jets observed in data. Overlaid are the fit
      projections of the various components.  The region between the
      vertical dashed lines is excluded from the fit.  (b)~The same
      distribution after subtraction of all SM components except the
      electroweak processes \PW\PW/\PW\cPZ. Error bars correspond to
      the statistical uncertainties.  The hatched band represents the
      uncertainty on the sum of the SM components including
      correlations from the fit.  The dark blue histogram is a
      resonance consistent with detector resolution and normalized to
      the CDF cross section scaled as described in the text.  (c)~The
      bin-by-bin pull, $(\text{data} - \text{fit})/(\text{fit
        uncertainty})$.  The bins in the figures are representative of
      the expected resolution for a given mass and the number of
      entries in each bin is scaled by its width.  }
    \label{fig:Fig1}
\end{figure*}

\begin{table*}[tbp]
  \topcaption{Event yields determined from maximum-likelihood fits to
    the data.  The total fit yields are corrected for bias.  The total
    fit uncertainties include the correlations among the various yields,
    as determined by the fit, and the corrections derived from the fit
    validation described in the
    text.  The $\chi^2$ probability uses the residuals and the data and MC
    statistical errors.}
  \label{tab:yields}
\scotchrule[lcccc]
                        & \multicolumn{2}{c}{muons}  & \multicolumn{2}{c}{electrons} \\
   Process              & 2-jet         & 3-jet       & 2-jet          & 3-jet \\
\hline
   \Wpj               & $58919 \pm 530$ & $13069 \pm 366$ & $29787 \pm 1153$ & $\phantom{0}8397 \pm 292$ \\
   Dibosons             & $\phantom{0}1236 \pm 114$  & $\phantom{00}333 \pm 32\phantom{0}$    & $\phantom{00}685 \pm 65\phantom{00}$     & $\phantom{00}184 \pm 18\phantom{0}$ \\
   \ttbar               & $\phantom{0}4570 \pm 307$  & $\phantom{0}9049 \pm 382$  & $\phantom{0}2556 \pm 174\phantom{0}$   & $\phantom{0}4265 \pm 253$ \\
   Single-top           & $\phantom{0}1765 \pm 87\phantom{0}$   & $\phantom{0}1001 \pm 50\phantom{0}$   & $\phantom{00}916 \pm 46\phantom{00}$     & $\phantom{00}521 \pm 26\phantom{0}$ \\
   Drell--Yan plus jets & $\phantom{0}1837 \pm 79\phantom{0}$   & $\phantom{00}561 \pm 24\phantom{0}$    & $\phantom{0}1061 \pm 46\phantom{00}$    & $\phantom{00}364 \pm 16\phantom{0}$ \\
   Multijet (QCD)       & $\phantom{000}29 \pm 284$    & $\phantom{0000}0 \pm 90\phantom{0}$      & $\phantom{0}3944 \pm 1133$  & $\phantom{00}324 \pm 160$ \\
\hline
  Fit $\chi^2$ probability&   0.454      & 0.729           & 0.969            & 0.991 \\
   Total from fit       & $68294 \pm 307$ & $24013 \pm 193$ & $38949 \pm 228\phantom{0}$  & $14055 \pm 143$ \\
   Data                 & 67900           & 24046           & 38973            & 14145 \\
\hline
\hline
\multicolumn{5}{c}{In the signal region $123 < m_{jj} < 186\GeV$ (excluded from the fit)} \\
\hline
   Total predicted      & $14511 \pm 125$ & $\phantom{0}7739 \pm 95\phantom{0}$   & $\phantom{0}7944 \pm 92\phantom{00}$    & $\phantom{0}4347 \pm 70\phantom{0}$ \\
   Data                 & 14050           & 7751            & 8023             & 4438 \\
\donescotchrule
\end{table*}

Figure~\ref{fig:Fig1}(a) shows the observed \mjj distribution for all
four event categories combined, together with the fitted projections
of the contributions of various SM processes.
Figure~\ref{fig:Fig1}(b) shows the same distribution after subtraction
of all SM contributions from data except electroweak diboson
\PW\PW/\PW\cPZ\ events.  No peak is visible in the spectrum except
that near 80\GeV due to diboson events.  Figure~\ref{fig:Fig1}(c)
shows the bin-by-bin pull.  Table~\ref{tab:yields} presents the
yields of the SM components obtained from the fit.  The sum of all
the contributions is compared to the number of observed events.  All
numbers except those in the last two rows are for the \mjj range of 40
to 400\GeV.  The last two rows compare the observed number of events
and the number predicted by the fit in the \mjj range of 123 to
186\GeV.  The data agree with the SM expectations, and we find no
significant excess in the signal region. We observe a sizable deficit
in the muon 2-jet data with respect to the prediction from our model.
We do not observe similar deviations in the other three categories,
suggesting it is a fluctuation and not a systematic bias.

We validate the fit procedure by performing pseudo-experiments.  In
each experiment, we generate the \mjj pseudo-data of the SM processes,
including the correlations taken from the fit to data, and then fit
each pseudo-data sample.  The results indicate that the bias on the
total yield is below 0.2\% and that the fit underestimates the total
yield uncertainty by about 30\%.  These effects are corrected for in
the final result.  Uncertainties in the jet energy are estimated using
a sample of \PW\ bosons decaying hadronically in a pure sample of
semileptonic \ttbar\ events.  The mean and resolution of the
reconstructed dijet mass distribution in data agree within 0.6\% with
the expectation from simulation.  A small difference in \met
resolution~\cite{Chatrchyan:2011tn} between data and simulation
affects the signal acceptance for the new physics models under
consideration at the 0.5\% level.  Further systematic uncertainties
are due to the uncertainty of the trigger efficiency estimates (1\%)
and the estimate of lepton reconstruction and selection efficiency
(2\%)~\cite{WZCMS:2010}.  The uncertainty on the integrated luminosity
is 2.2\%~\cite{lumiPAS}.

We scrutinize the dijet mass spectrum near 150\GeV, searching for a
technicolor, leptophobic $\zp$, or \PW\PH\ resonant enhancement.  We
also use a generic signal model obtained by convolving a delta
function centered at $\mjj=150\GeV$ with a Gaussian function having
width equal to $\sigma_{jj}$.  Figure~\ref{fig:Fig1}(b) shows this
generic signal shape.  The expected number of signal events at the LHC
for a given cross section at the Tevatron can be estimated by
considering the ratio of the predicted cross sections for our
reference process, \PW\PH\ production with $M_{\text{H}} = 150\GeV$.
This process is dominated by quark-antiquark (\cPq\cPaq) annihilation.
As \cPq\cPaq\ processes have the smallest increase in parton
luminosity from the Tevatron to the LHC, this choice provides a
conservative limit.  We therefore assume
\begin{linenomath}
\begin{equation}
\sigma_{\text{LHC}}^{\text{dijet resonance}} =
\sigma_{\text{Tevatron}}^{\text{dijet resonance}}
\frac{\sigma_{\text{LHC}}^{\PW\PH}}{\sigma_{\text{Tevatron}}^{\PW\PH}},\label{eqTevToLHC2}\nonumber
\end{equation}
\end{linenomath}
where $\sigma_{\text{LHC}}^{\PW\PH} =
300.1\unit{fb}$~\cite{LHC4PDFxsec} and
$\sigma_{\text{Tevatron}}^{\PW\PH} =
71.8\unit{fb}$~\cite{Carena:2000yx}.  A generic Gaussian signal
normalized to $\sigma_{\text{Tevatron}} = 4\unit{pb}$ corresponds to
$\sigma_{\text{LHC}} = 16.7\unit{pb}$.  Table~\ref{tab:signals}
contains the values of $\sigma_{\text{LHC}}$ times the branching
fraction to jets and of the overall efficiency times acceptance
$\varepsilon\mathcal{A}$ for the models considered.

\begin{table}[btp]
  \topcaption{\label{tab:signals} The \PYTHIA cross
    sections at 7\TeV times branching fraction to jets
    ($\sigma\times\mathcal{B}$) and
    overall efficiency times acceptance ($\varepsilon\mathcal{A}$) for
    various signal models.  The relative uncertainties in $\varepsilon$
    measurements are 1--2\%.  The uncertainty on $\mathcal{A}$ is negligible.}
\scotchrule[{l c c c c c}]
    &   & \multicolumn{4}{c}{$\varepsilon\mathcal{A}$} \\
    &   & \multicolumn{2}{c}{muons} & \multicolumn{2}{c}{electrons} \\
    Signal model &  $\sigma\times\mathcal{B}$ (pb) & 2-jet & 3-jet & 2-jet & 3-jet \\
    \hline
    Technicolor~\cite{Eichten:2011sh} & 7.4   & 0.065 & 0.020 & 0.039 & 0.011 \\
    $\zp$~\cite{Buckley:2011vc}       & 8.1   & 0.070 & 0.023 & 0.042 & 0.014 \\
    \PW\PH~\cite{Sjostrand:2006za}    & 0.059 & 0.060 & 0.019 & 0.038 & 0.013 \\
\donescotchrule
\end{table}

\begin{figure}[btp]
    \begin{center}
      \includegraphics[width=\columnfigure]{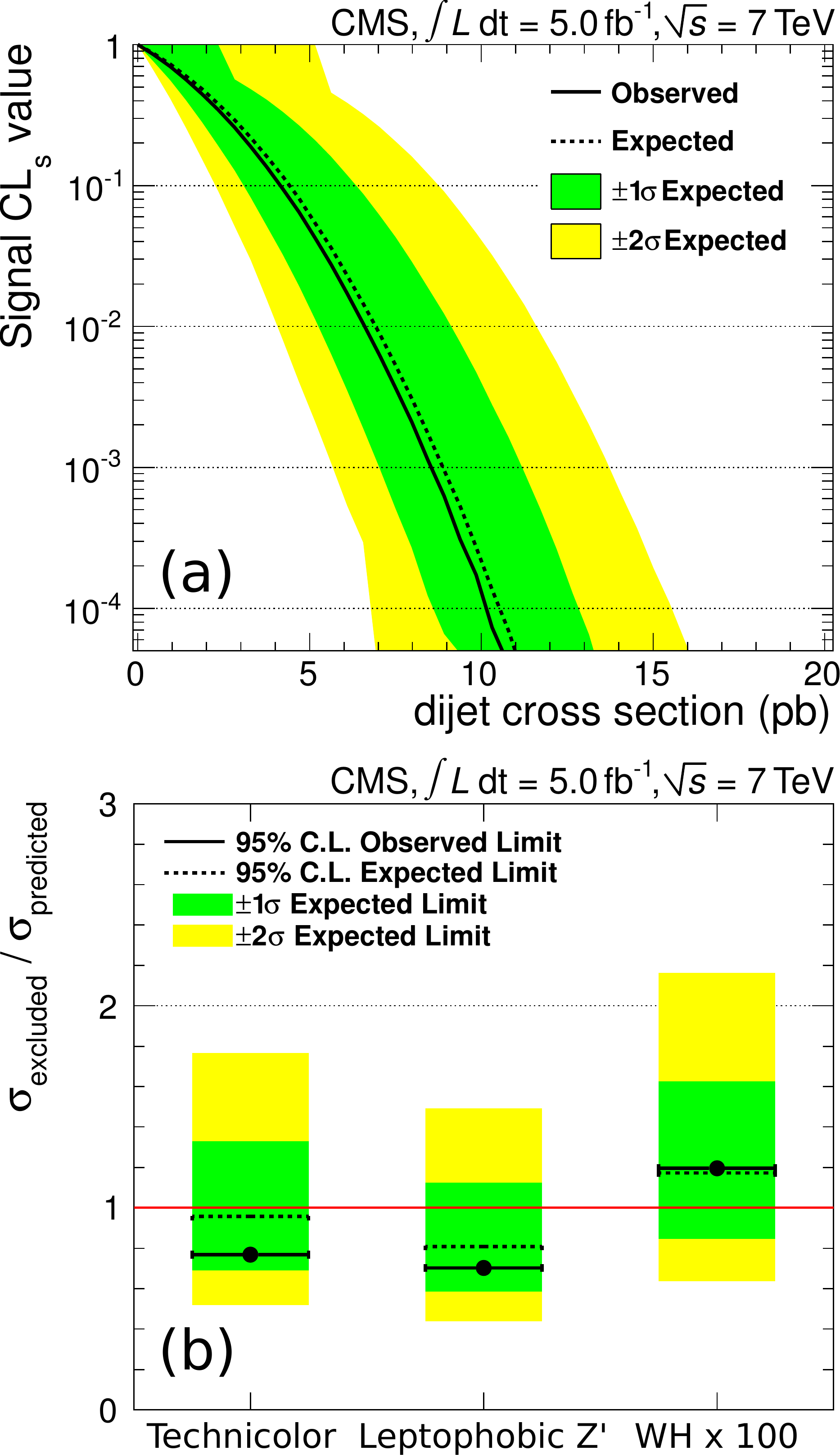}
    \end{center}
    \caption{\label{fig:Fig2}(a) The observed and expected values of
      the $\text{CL}_{\text{S}}$ statistic for a generic Gaussian
      signal hypothesis with $M=150\GeV$ and $\sigma_{jj}=15\GeV$, as a
      function of the dijet signal cross section.  (b)~Observed and
      expected 95\% CL upper limits, with one- and two-sigma error
      bands, on the cross section divided by the expected values for
      various signal models.  The limits are calculated using the
      $\text{CL}_{\text{S}}$ method. A value of the excluded cross
      section over the predicted cross section of less than one
      indicates that the model is excluded at 95\% CL.
      Table~\ref{tab:signals} lists the cross sections for these
      models.  }
\end{figure}

Since we observe no resonant enhancement, we proceed to set exclusion
limits using a modified frequentist $\text{CL}_{\mathrm{S}}$
method~\cite{CLS,Junk:1999kv} with profile likelihood as the test
statistic.  Inputs to the limit-setting procedure are the \mjj
distribution obtained by combining the SM components from the fit, the
observed distribution in data, the expectation from the dijet
resonance model under consideration and the uncertainties associated
with these quantities.  Figure~\ref{fig:Fig2}(a) shows the observed
and expected $\text{CL}_{\text{S}}$ values versus cross section for a
generic Gaussian signal, after combining the results of all four event
categories.  We set a 95\% CL upper limit of 5.0\unit{pb} and a 99.9\%
CL upper limit of 8.5\unit{pb} on the dijet production cross section
for a generic resonance with \PW\PH-like $\varepsilon\mathcal{A}$.

Figure~\ref{fig:Fig2}(b) compares the 95\% CL upper limits with the
expected cross sections for technicolor, leptophobic $\zp$, and \PW\PH
($M_{\text{H}} = 150\GeV$) signals.  The technicolor and $\zp$ models
are excluded. Because we have minimal sensitivity to \PW\PH, we
compare the limit in Fig.~\ref{fig:Fig2}(b) to 100 times the SM cross
section as an illustration.

In summary, we have studied the invariant mass spectrum of the two
jets with highest transverse momentum in
$\text{pp}\rightarrow\PW$+2-jet and \PW+3-jet events, with the \PW\
decaying leptonically to a muon or electron.  The analyzed data sample
corresponds to an integrated luminosity of 5.0\fbinv at $\sqrt{s} =
7\TeV$.  We find no evidence for a resonant enhancement near a dijet
mass of 150\GeV, as reported by the CDF Collaboration, and set upper
limits on the dijet production cross section of 5.0\unit{pb} at 95\%
CL and 8.5\unit{pb} at 99.9\% CL.  Two theoretical models, leptophobic
$\zp$ and technicolor, which predict the presence of a resonant
enhancement near 150\GeV, are excluded.

We thank Adam Martin and Matthew Buckley for help with simulation of
technicolor and \zp models, respectively.  We congratulate our
colleagues in the CERN accelerator departments for the excellent
performance of the LHC machine. We thank the technical and
administrative staff at CERN and other CMS institutes, and acknowledge
support from FMSR (Austria); FNRS and FWO (Belgium); CNPq, CAPES,
FAPERJ, and FAPESP (Brazil); MES (Bulgaria); CERN; CAS, MoST, and NSFC
(China); COLCIENCIAS (Colombia); MSES (Croatia); RPF (Cyprus); MoER,
SF0690030s09 and ERDF (Estonia); Academy of Finland, MEC, and HIP
(Finland); CEA and CNRS/IN2P3 (France); BMBF, DFG, and HGF (Germany);
GSRT (Greece); OTKA and NKTH (Hungary); DAE and DST (India); IPM
(Iran); SFI (Ireland); INFN (Italy); NRF and WCU (Korea); LAS
(Lithuania); CINVESTAV, CONACYT, SEP, and UASLP-FAI (Mexico); MSI (New
Zealand); PAEC (Pakistan); MSHE and NSC (Poland); FCT (Portugal); JINR
(Armenia, Belarus, Georgia, Ukraine, Uzbekistan); MON, RosAtom, RAS
and RFBR (Russia); MSTD (Serbia); MICINN and CPAN (Spain); Swiss
Funding Agencies (Switzerland); NSC (Taipei); TUBITAK and TAEK
(Turkey); STFC (United Kingdom); DOE and NSF (USA).

\bibliography{auto_generated}   

\cleardoublepage \appendix\section{The CMS Collaboration \label{app:collab}}\begin{sloppypar}\hyphenpenalty=5000\widowpenalty=500\clubpenalty=5000\input{EWK-11-017-authorlist.tex}\end{sloppypar}
\end{document}

%% file: EWK-11-017-authorlist.tex
\textbf{Yerevan Physics Institute,  Yerevan,  Armenia}\\*[0pt]
S.~Chatrchyan, V.~Khachatryan, A.M.~Sirunyan, A.~Tumasyan
\vskip\cmsinstskip
\textbf{Institut f\"{u}r Hochenergiephysik der OeAW,  Wien,  Austria}\\*[0pt]
W.~Adam, E.~Aguilo, T.~Bergauer, M.~Dragicevic, J.~Er\"{o}, C.~Fabjan\cmsAuthorMark{1}, M.~Friedl, R.~Fr\"{u}hwirth\cmsAuthorMark{1}, V.M.~Ghete, J.~Hammer, N.~H\"{o}rmann, J.~Hrubec, M.~Jeitler\cmsAuthorMark{1}, W.~Kiesenhofer, V.~Kn\"{u}nz, M.~Krammer\cmsAuthorMark{1}, I.~Kr\"{a}tschmer, D.~Liko, I.~Mikulec, M.~Pernicka$^{\textrm{\dag}}$, B.~Rahbaran, C.~Rohringer, H.~Rohringer, R.~Sch\"{o}fbeck, J.~Strauss, A.~Taurok, W.~Waltenberger, G.~Walzel, E.~Widl, C.-E.~Wulz\cmsAuthorMark{1}
% ...
\vskip\cmsinstskip
\dag:~Deceased\\
1:~~Also at Vienna University of Technology, Vienna, Austria\\
% ...
